# An HAM-Based Analytic Modeling Methodology for Memristor Enabling Fast Convergence

Wei Hu, Du Yongqian, Haibo Luo, Chuandong Chen, and Rongshan Wei, *Member, IEEE*

*Abstract*—Memristor has great application prospects in various high-performance electronic systems, such as memory, artificial intelligence, and neural networks, due to its fast speed, nano-scale dimensions, and low-power consumption. However, traditional nonanalytic models and lately reported analytic models for memristor have the problems of nonconvergence and slow convergence, respectively. These problems lay great obstacles in the analysis, simulation and design of memristor. To address these problems, a modeling methodology for analytic approximate solution of the state variable in memristor is proposed in this work. This methodology solves the governing equation of memristor by adopting the Homotopy Analysis Method (HAM) for the first time, and the convergence performance of the methodology is enhanced by an optimized convergence-control parameter $\hbar$ in the HAM. The simulation results, compared with the reported analytic models, demonstrate that the HAM-based modeling methodology achieves faster convergence while guaranteeing sufficient accuracy. Based on the methodology, it can be simultaneously revealed that highly nonlinearity and large $\xi$ are the potential sources of slow convergence in analytic models, which is beneficial for the analysis and design guidance. In addition, a Spice subcircuit is constructed based on the obtained HAM model, and then it is integrated into an oscillator to verify its applicability. Due to the generality of HAM, this methodology may be easily extended to other memory devices.

## I. INTRODUCTION

MEMRISTOR (also known as Resistive Random Access Memory) can be used in memory [1]–[4], neurobiology [5]–[7], artificial intelligence [8]–[10], and neural networks [11], [12], etc., with great application prospects, due to its advantages such as fast speed [13], nanometers scale [14], and low-power consumption [15], [16]. Memristor is characterized by nonvolatile memory effect, whose resistance is related to the history of excitation signal and internal state variable. Since the successful development of the Hewlett-Packard (HP) prototype device in 2008 [17], memristor has gained a great deal of research interests [18].

A well established model is extremely crucial for the analysis and simulation of memristor, and it is also a precondition for memristor-based integrated circuit (IC) designing. However, previously works on memristor modeling methodology mainly focus on nonanalytic models (e.g., the microscopic physical models [19], [20]), which may result in nonconvergence [21]. Some other methodologies adopt numerical integration method (e.g., the Finite Elements method and the Runge–Kutta method), which may cause truncation and rounding errors while approximating the solution of the state variable [22]. As a result, once the models based on these methodologies are embedded in electronic systems, nonconvergence may occur, particularly in large-scale circuits [23].

To address the above nonconvergence problems, following modeling methodologies are adopted: 1) Compact modeling methodology [24]–[26]. It exploits the basic circuit elements in Spice subcircuit to simulate the dynamic evolutions of the state variables of governing equations (also called the nonlinear differential equations of the state variable) in memristors. The models based on this methodology offer fast convergence but have limited scalability and weak physical significance; 2) Phenomenological modeling methodology [27], [28]. This methodology, which successfully solves the nonconvergence problem, is strictly analytic because the fitting strategy, instead of the numerical integration method, is adopted to simulate the state variable. Unfortunately, its scalability is not enough, due to the fact that the models based on this methodology are only customized for nanodevices with specific physical structures and mechanisms; 3) Homotopy Perturbation Methodology (HPM) [29]. It is an effective methodology for approximating the analytic solution with the homotopy-series, and it was firstly proposed by He [30] in 2000. In 2017, HPM was adopted by Hernández-Mejía et al. [29] to solve the governing equation of the HP memristor. They have established a semi-symbolic approximate analytic methodology with the practical physical parameters, which effectively overcomes the nonconvergence caused by the nonanalytic models. However, because this methodology is not optimized for the homotopy-series solution, the convergence speed of solution is slow and uncontrollable. In worst case, nonconvergence still happens. (These limitations will be analyzed in Section IV in detail.) As a result, the high approximation order, which results in complicated expression, is needed to meet the specific accuracy requirements because of the slow convergence.

Inspired by the HPM methodology [29], we propose a novel memristor modeling methodology in this work to overcome the above limitations while maintaining the analyticity and having high scalability. *Homotopy Analysis Method* (HAM) [31]–[33], which has been widely used in mathematics community, is employed for the first time in the modeling of memristor to solve the governing equation and accelerate convergence. With the help of the proposed HAM-based methodology, we reveal the potential sources of the slow convergence in analytic model. To the best of the authors' knowledge, no available HAM based methodology accounts for the memristor modeling.

The structure of this paper is organized as follows: Section II reviews the related works on HP memristor and HAM. Section III proposes an analytic HAM-based modeling methodology. Section IV verifies the obtained models and compares them with the exact solution based and HPM based models. Section V builds a HAM-based memristor Spice subcircuit and then its applicability is verified by an oscillator. Section VI, followed by the conclusion in Section VII, is devoted to comparing the methodology with others in various aspects, including thorough discussions of its characteristics.

## II. FUNDAMENTALS OF MEMRISTOR AND HAM

### A. HP Memristor

Because of its milestone significance, the HP memristor [17] was employed in this work as an example model to illustrate the proposed modeling methodology. Fig. 1 shows the structure of the HP memristor [17]. It is a *Pt/TiO₂/Pt* sandwich structure, which contains two parts: one is the *platinum* (*Pt*) electrodes at both terminals, and the other is the thin *titanium dioxide* film ($TiO_2$ switching layer) with doped oxygen ions sandwiched between the two *Pt* electrodes. The oxygen ions drift nonlinearly in the switching layer under a strong internal electric field introduced by an external excitation signal (voltage or current), changing ratio (state variable) between the doped layer *w* and the total thickness *D*. As a result, the *memristance-modulation* in real time is realized. In other words, the memristance is changed depending on the amplitude, polarity, and duration of the excitation signal.

The governing equation, port equation, and memristance of the HP memristor model can be expressed respectively by [17]

$$\frac{dx(t)}{dt} = \frac{\mu_V R_{ON}}{D^2} I_M(t) f[x(t)] \quad (1)$$
$$V_M(t,x) = R_M(t,x) I_M(t) \quad (2)$$
$$R_M(t,x) = R_{ON} x(t) + R_{OFF}[1-x(t)] \quad (3)$$

where normalized state variable $x(t) = w(t)/D \in [0,1]$ [which is the core of the memristor that determines the other physical parameters, such as output voltage (2) and memristance (3)], input current $I_M(t) = A\sin(\omega t)$. A window function $f[x(t)]$ is introduced in (1), allowing the HP model to ensure $x \in [0,1]$ and to overcome the boundary effect [34]. Even the commonly used Joglekar window functions (J-windows) [35] were chosen in this work, the relevant analyses and conclusions could also be applied to other important windows, such as Biolek [34], Prodromakis [36], and VTEAM windows [37].

Note the governing equation (1) is a nonlinear differential equation that controls the evolution of the state variable, which is suitable to be solved by HAM because HAM is commonly used to solve various nonlinear differential equations [33]. Modeling a memristor is essentially a procedure of obtaining the state variable by governing equation. All the parameters of the HP memristor and the excitation are summarized in Table I.

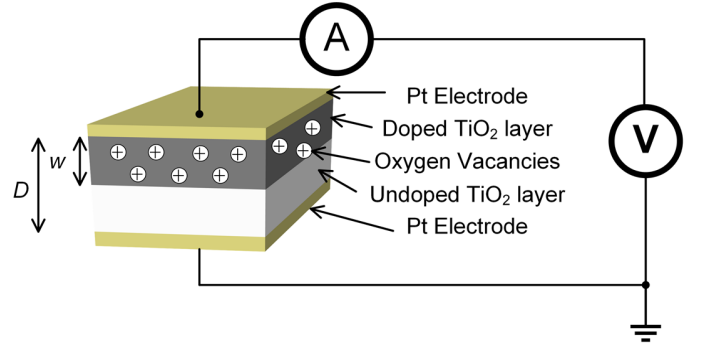

Fig. 1. Device structure of the HP memristor [17].

TABLE I
PHYSICAL PARAMETERS AND EXCITATION OF HP MEMRISTOR

| Parameter | Value | Unit | Definitions |
|---|---|---|---|
| $D$[a] | $10\times10^{-9}$ | m | Thickness of the switching layer |
| $w$[a] | $[0, 10^{-8}]$ | m | Length of the doped layer |
| $R_{OFF}$[a] | $16\times10^3$ | Ω | High resistance ($x$=0) |
| $R_{ON}$[a] | $10^2$ | Ω | Low resistance ($x$=1) |
| $\mu_V$[a] | $10^{-14}$ | m²s⁻¹V⁻¹ | Average ion mobility in small electric field |
| $x(t)$ | [0,1] | -- | Normalized state variable ($w(t)/D$) |
| $x_0$ | $10^{-1}$ | -- | Initial state variable when $t=0$ |
| $\omega$ | 1 | rad/s | Angular frequency of sinusoidal excitation signal |
| $A$ | $40\times10^{-6}$ | A | Magnitude of sinusoidal excitation signal |
| $V_M$ | $[-\infty, \infty]$ | V | Voltage across the memristor |
| $I_M$ | $[-\infty, \infty]$ | A | Current through the memristor |
| $R_M$ | $[0, \infty]$ | Ω | Memristance |

[a] The parameters are obtained from [17].

### B. Homotopy Analysis Method

In HAM, the nonlinear differential equation, like (1), has a general expression

$$\mathcal{N}[u(t)] = 0 \quad (4)$$

where $\mathcal{N}$ is a nonlinear operator, $u(t)$ is an unknown function need to be solved, and $t$ is a time variable.

To analytically solve (4), the following *homotopy zero-order deformation equation* [38] is constructed:

$$(1-q)\mathcal{L}[\phi(t;q) - u_0(t)] + \hbar q H(t)\mathcal{N}[\phi(t;q)] = 0 \quad (5)$$

where $\phi(t;q)$, used to approximate the unknown $u(t)$, is an unknown function of an embedded parameter $q \in [0,1]$, $\hbar \neq 0$ is an introduced convergence-control parameter, $H(t) \neq 0$ is an auxiliary function, $\mathcal{L}$ is a linear operator, and $u_0(t)$ is an initial guess of $u(t)$. From (5), we can see $\phi(t;q)$ and $u(t)$ are linked by homotopy which is a mathematical theory that facilitates solving a nonlinear equation by converting it into several linear equations [38].

When substitute $q = 0$ and $q = 1$ into (5), the following equations hold:

$$\phi(t;0) = u_0(t) \quad (6)$$
$$\phi(t;1) = u(t). \quad (7)$$

For this reason, $\phi(t;q)$ transforms from the initial guess $u_0(t)$ to $u(t)$ as $q$ increases from 0 to 1, which is a typical characteristic of the homotopy [32].

Assuming that $\phi(t;q)$ is a continuous smooth function with respect to $q$, then it is can be expanded into a Maclaurin-series in terms of $q$

$$\phi(t;q) = u_0(t) + \sum_{m=1}^{+\infty} u_m(t) q^m \quad (8)$$

where $m$ is the number of iterations, $u_m(t)$ is the *mth-order homotopy coefficient* and $\mathscr{D}_m$ is the *mth-order homotopy derivative operator*, which can be expressed as

$$u_m(t) = \mathscr{D}_m[\phi(t;q)] \quad (9)$$
$$\mathscr{D}_m = \frac{1}{m!} \frac{\partial^m}{\partial q^m} \Big|_{q=0}. \quad (10)$$

When $q = 1$

$$\phi(t;1) = u(t) = u_0(t) + \sum_{m=1}^{+\infty} u_m(t) \quad (11)$$

which is one of the solutions of the nonlinear differential equation (4).

To solve (11), differentiating (5) $m$ times in term of $q$ and letting $q = 0$, then dividing both sides of the them by $m!$, the *high-order deformation equation* is obtained

$$\mathscr{L}\big[u_m(t) - \chi_m u_{m-1}(t)\big] = \hbar H(t) \mathscr{D}_m(\vec{u}_{m-1}) \quad (12)$$

where

$$\chi_m = \begin{cases} 0, & m \leq 1 \\ 1, & m > 1 \end{cases} \quad (13)$$

and $\vec{u}_m$ is a vector defined as

$$\vec{u}_m = \{u_0(t), u_1(t), \ldots, u_m(t)\}. \quad (14)$$

Combining (7), (11), and (12), the analytic homotopy-series solution of $u(t)$ can be deduced as

$$u(t, \hbar) = u_0(t) + \sum_{m=1}^{N} u_m(t, \hbar) \quad (15)$$

where $N$ is the approximation order and $u_m(t, \hbar)$ can be solved from (12).

It should be noted that the convergence-control parameters $\hbar$ drastically affects the convergence of (15). As a result, solving the optimal $\hbar$ is a crucial step in HAM to obtain a complete expression of (15) with the fastest convergence. The optimal $\hbar$ can be solved by minimizing $Em(\hbar)$

$$\frac{dEm(\hbar)}{d\hbar} = 0 \quad (16)$$

where $Em$ is the *discrete squared residual error* [39] of the nonlinear differential equation (4), namely

$$Em(\hbar) \approx \frac{1}{N_P + 1} \sum_{j=0}^{N_P} \left\{ \mathscr{N}\left[ \sum_{m=0}^{N} u_m(t_j, \hbar) \right] \right\}^2. \quad (17)$$

In (17), $N_P$ is the number of discrete points with $t_j = j\Delta t$, and

$$\Delta t = \frac{\Gamma}{N_p} \quad (18)$$

where $\Gamma$ is the effective range of $t$ for (4) under consideration.

Theoretically, the more quickly $\hbar$ approaches its optimal value (which corresponds to the minimum $Em$), the faster $u(t, \hbar)$ converges to the exact solution of (4).

Finally, by substituting the optimized $\hbar$ (16) into (15), and then solving the linear deformation equations, we obtain the $N$ th-order analytic approximate solution (homotopy-series solution)

$$u_N(t) = u_0(t) + \sum_{m=1}^{N} u_m(t). \quad (19)$$

According to (19), we could have an exact solution if $N \to \infty$.

## III. METHODOLOGY DESCRIPTION

In this section, a modeling methodology for the state variable of memristor is presented based on the HAM introduced in Section II. The detail of the methodology is summarized in Algorithm 1, where the symbols and operators used in this methodology are summarized in Table II. Then, we apply the Algorithm 1 to the physical model of HP memristor for solving the analytic approximate solution of the state variable $x(t)$. Base on the solved $x(t)$, we obtain the approximate analytic HP model.

Depending on the Algorithm 1 and the characteristics of the governing equation (1) of the original HP memristor model, we firstly define $x(t) = u(t)$ and then give the corresponding nonlinear operator

$$\mathscr{N}[x(t)] = \mu_{\text{V}} \frac{R_{ON} A}{D^2} \sin(\omega t) f[\phi(t;q)]. \quad (20)$$

The associated linear operator can be chosen as

$$\mathscr{L}[\phi(t;q)] = \frac{\partial \phi(t;q)}{\partial t} \quad (21)$$

with the property

$$\mathscr{L}[C] = 0 \quad (22)$$

where $C$ is a constant. For the sake of simplicity, we define the auxiliary function $H(t) = 1$. Note that the determination of the linear operator mainly depends on the characteristics of the governing equation, and the auxiliary function is selected based on the *rule of solution expression* and *regularity of solutions* [38].

After performing inverse linear operator $\mathscr{L}^{-1}$ [based on (21)] on both sides of (12) and letting $x(t) = u(t)$, following equation is reached:

$$\begin{aligned}\mathscr{L}^{-1}\left[\mathscr{L}[x_m(t) - \chi_m x_{m-1}(t)]\right] \\= \mathscr{L}^{-1}[\hbar H(t)\mathscr{D}_m(\vec{x}_{m-1})].\end{aligned} \quad (23)$$

Combining (9), (20), and (23), and rearranging the related equations, the solution of the $m$th-order deformation equation can be expressed as

$$\begin{aligned}x_m(t,\hbar) = \hbar \int_0^t H(\tau)\mathscr{D}_{m-1}(\vec{x}_{m-1})\,d\tau \\+ \chi_m x_{m-1}(t) + C\end{aligned} \quad (24)$$

where $C$ can be derived from $x_0$ and (22).

Finally, after solving $\hbar$ with (16) and substituting (24) into (19), the $N$th-order analytic approximate solution (also called the homotopy-series solution) of the state variable $x(t)$ can be deduced as

$$x_N(t) = x_0(t) + \sum_{m=1}^{N} x_m(t). \quad (25)$$

For simplicity, this solution is named as HAM$N$.

These systematic descriptions of the proposed methodology are preconditions for a clear understanding of verifications and potential sources of slow convergence problems, which are presented in the next section.

## IV. Methodology Verification

For completeness, the HAM-based methodology was verified by comparing simulations of the obtained models and relevant HPM models [29] with exact and numerical solutions based models in three cases: 1) When the model has an exact solution (weakly nonlinearity); 2) Highly nonlinearity; and 3) Large $\xi$. Note all numerical simulations were performed by using the fourth-order Runge–Kutta method, while analytic simulations were implemented by using Mathematica. Then the potential sources of the slow convergence in analytic model were revealed according to these comparisons.

Two special emphases before the verifications are listed as follows:
1) It was assumed that the excitation in each verification is a current signal $A\sin(\omega t)$. Nevertheless, similar verifications could also be applied if a voltage signal is used [34], [37].
2) $N = 3$ and $N = 6$ were chosen for all case studies.

Adopting the approximate analytic HP model in Section III, we verified the HAM-based modeling methodology in the following three cases determined by different J-windows [35]:

### A. Exact solution (weakly nonlinearity)

The HP model has an exact solution of the state variable $x(t)$ when using the J-window with $P=1$ ($P$ is a control parameter

---

**Algorithm 1:** Analytic HAM-based modeling methodology for memristor
**Inputs**: Prototypical physical model, device parameters, and current excitation $I_M(t)$[a].
**Outputs**: State variable $x(t)$, memristance $R_M(x)$, voltage $V_M(x)$.
**Event**: Using homotopy-series to approximate the analytical solution of $x(t)$ by solving the governing equation, and introducing the convergence-control parameter $\hbar$ to control the convergence of the solved $x(t)$, then the analytical solutions of the outputs based on $x(t)$ are obtained.

1. Determine the approximation order $N$;
2. Construct the zero-order deformation equation (5) and the $m$th-order deformation equation (12) of $x(t)$ based on the governing equation (1);
3. Choose the auxiliary function $H(t)$, the linear operator $\mathscr{L}$ and the initial approximate solution $x_0$ in (5);
4. **for** $m = 1$, **do**
5.    Solve $m$th-order deformation equation;
6.    $m = m + 1$;
7.    **if** $m > N$, **then**
8.       Stop the iteration, let $x_N(t,\hbar) = u_N(t,\hbar)$;
9.    **end if**
10. **end for**
11. $\frac{dEm(\hbar)}{d\hbar} = 0 \to \hbar$ (16);
12. $x_N(t,\hbar) \xrightarrow{\hbar} x_N(t) = x_0(t) + \sum_{m=1}^{N} x_m(t)$;
13. **return** $x_N(t) = x_0$

[a] This algorithm is also applicable to voltage excitation by transposition of current and voltage in the port equation (2).

TABLE II
SYMBOLS AND OPERATORS USED FOR HAM-BASED MEMRISTOR MODELING METHODOLOGY

| Symbols and operators | Definitions | Descriptions |
|---|---|---|
| $\mathscr{N}$ | Nonlinear operator | Presents a nonlinear system |
| $\mathscr{L}$ | Linear operator | Constructs the deformation equation |
| $\mathscr{D}_m$ | $m$th-order deformation derivative operator | Constructs the deformation equation |
| $H(t)$ | Auxiliary function | Constructs the deformation equation |
| $E_m$ | Discrete squared residual error | Its minimum determines $\hbar$ |
| $\hbar$ | Convergence-control parameter | Controls the convergence of solutions, $\hbar \in \mathbb{R}$ |
| $u_m(t)$ | $m$th-order homotopy-series | Constitutes the analytic approximate solution of the state variable $x(t)$ |

for nonlinearity and a positive integer. When $P=1$, that means the governing equation is weakly nonlinear) [35]. The exact solution was chosen as a reference for comparisons in this section. The associated governing equation take the form

$$\frac{dx(t)}{dt} = 4\mu_V \frac{R_{ON}A}{D^2}\sin(\omega t)(-x^2 + x). \quad (26)$$

with an exact solution

$$x(t) = \frac{x_0}{(1-x_0)\exp[4\xi(\cos(\omega t) - 1)] + x_0} \quad (27)$$

where

$$\xi = \frac{\mu_V R_{ON} A}{D^2 \omega}. \quad (28)$$

Note $\xi$ combines the physical parameters of the device and the input, its effect on convergence is analyzed in Section IV-C.

Implementing the Algorithm 1 in Section III and related arrangements, following third-order homotopy-series solution HAM3 is obtained for comparison of the exact solution (27):

$$x_3(t,\hbar) = x_0 + \sum_{m=1}^{3} x_m(t,\hbar) =$$
$$\begin{bmatrix} x_0 \\ +\hbar(\hbar^2 + 3\hbar + 3)p_{x_3,1} \\ +3(2\hbar + 3)\hbar^2 p_{x_3,2} \\ +10\hbar^3 p_{x_3,3} \end{bmatrix}$$
$$+ \cos(\omega t) \begin{bmatrix} -\hbar(\hbar^2 + 3\hbar + 3)p_{x_3,1} \\ -4(2\hbar + 3)\hbar^2 p_{x_3,2} - 15\hbar^3 p_{x_3,3} \end{bmatrix} \quad (29)$$
$$+ \cos(2\omega t) \begin{bmatrix} (2\hbar + 3)\hbar^2 p_{x_3,2} \\ +6\hbar^3 p_{x_3,3} \end{bmatrix}$$
$$- \cos(3\omega t)\,\hbar^3 p_{x_3,3}$$

where $p_{x_3,1}$, $p_{x_3,2}$, and $p_{x_3,3}$ are variables for expressions simplification, which are defined as

$$p_{x_3,1} = 4x_0(x_0 - 1)\xi; \quad (30)$$
$$p_{x_3,2} = 4x_0(x_0 - 1)(2x_0 - 1)\xi^2; \quad (31)$$
$$p_{x_3,3} = 8x_0(x_0 - 1)(6x_0^2 - 6x_0 + 1)\xi^3/3. \quad (32)$$

It can be seen from (29) that the analytic approximate solution is controlled by $\hbar$. The optimal $\hbar = -0.64172$ for the fastest convergence of (29) is derived from (16) and listed in Table III. After substituting it into (29), complete $x_3(t)$ is obtained. Note we only list $x_3(t,\hbar)$ in this paper due to the limited space, while other approximate order based solutions have the similar expressions.

To prove the validity and fast convergence of the model produced by the HAM-based methodology, we performed the following verifications:

TABLE III
MINIMUM $E_m$ AND OPTIMAL $\hbar$ FOR THE HAM AND HPM MODELS USING DIFFERENT WINDOW FUNCTIONS

| Windows | J-window [35] with $P = 1$ | | J-window [35] with $P = 2$ | |
|---|---|---|---|---|
| Third-order approximation | HPM3 | **HAM3** | HPM3 | **HAM3** |
| $E_m$ [a] | 22.4199 | **0.94683** | 151.4452 | **1.97935** |
| $\hbar$ [b] | -- | **-0.64172** | -- | **-0.43949** |
| Sixth-order approximation | HPM6 | **HAM6** | HPM6 | **HAM6** |
| $E_m$ [a] | 4.79866 | **0.13856** | 803.88821 | **2.69340 × 10⁻⁴** |
| $\hbar$ [b] | -- | **-0.87148** | -- | **-0.35493** |

[a] The $Em(\hbar)$ is derived from (17) and (18), the effective interval of $\Gamma$ is [0,1000], $N_P = 2000$.
[b] The optimal convergence-control parameter $\hbar$ is obtained from (16).

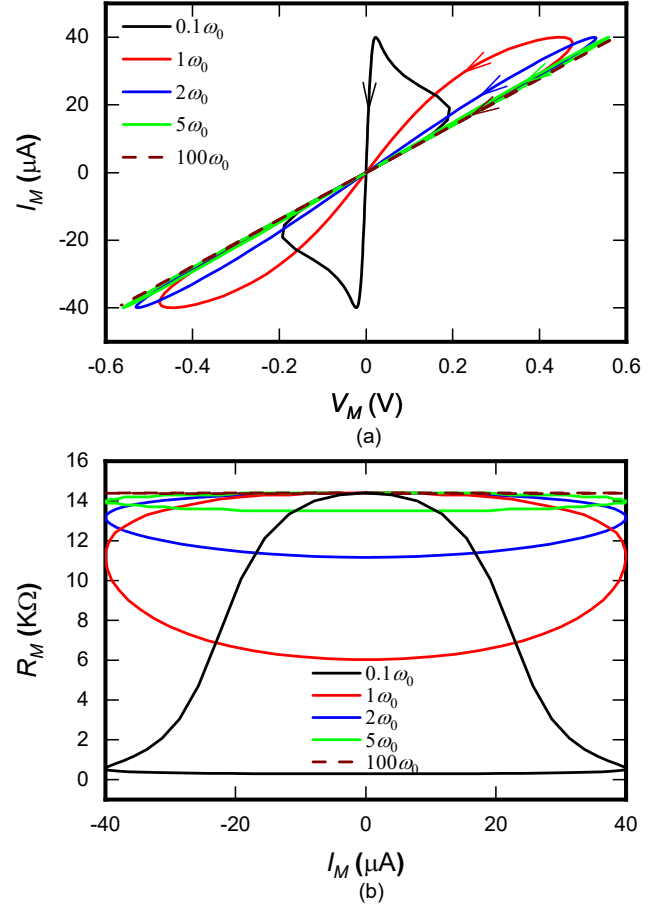

Fig. 2. Validity verifications of the obtained HAM3-based HP model with the classical fingerprints of memristor [40]. (a), (b) Simulated I-V and memristance curves under different sinusoidal inputs with increasing angular frequency. $\hbar$ is given in Table III. The simulation parameters of the memristor and excitation are given in Table I except: $\omega_0 = 1$ rad/s. Note the HAM3 is the third-order analytic approximate solution.

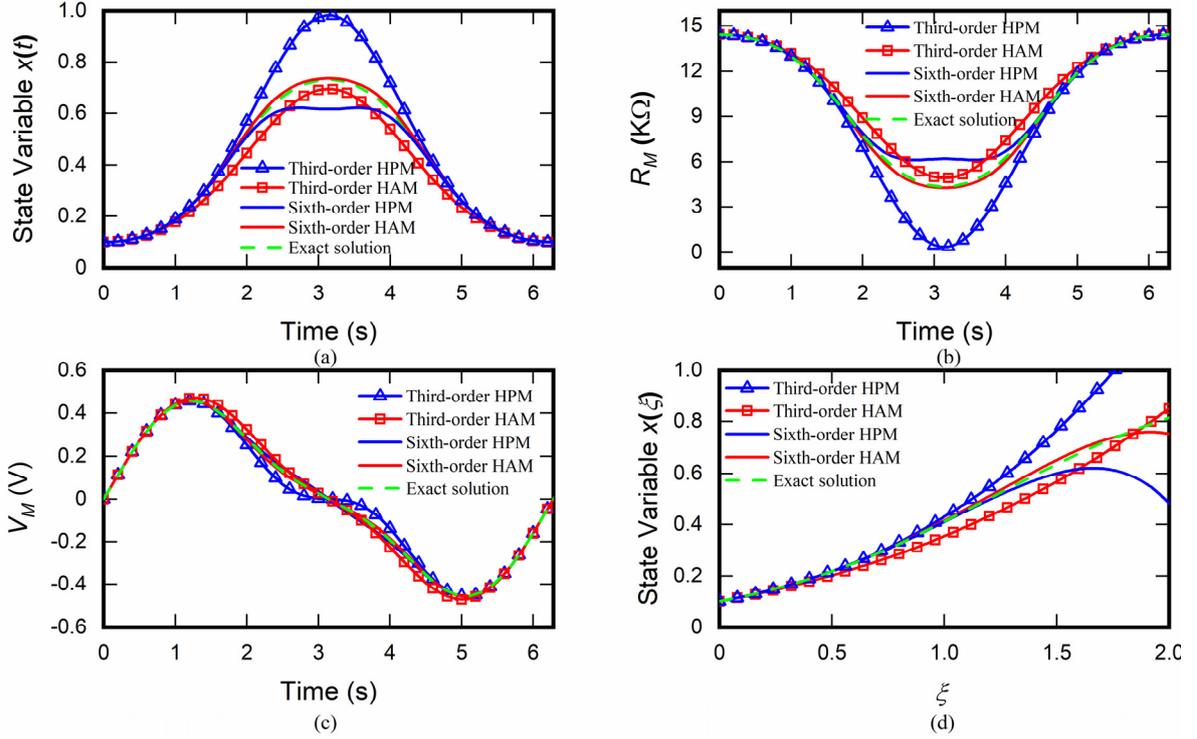

Fig. 3. Comparisons of the obtained HAM model and the HPM model [29] with the exact solution (27) based model. (a)–(c) the evolution of $x(t)$, $R_M$, and $V_M$, respectively. (d) the $x-\xi$ curve when $t=1$. The J-window [35] with $P=2$ is adopted. Approximation order $N=3$ and $N=6$, respectively. $\hbar$ is given in Table III. All the other simulation parameters of the memristor and excitation are given in Table I.

TABLE IV
CONVERGENCE ERROR COMPARISONS OF THE HAM AND HPM MODELS WITH THE EXACT SOLUTION BASED MODEL WHEN USING J-WINDOW WITH $P=1$

| Key physical parameters | State variable $x(t)$ | | $R_M(t)$ | | $V_M(t)$ | | State variable $x(\xi)$[b] | |
|---|---|---|---|---|---|---|---|---|
| Third-order approximation | HPM3 | **HAM3** | HPM3 | **HAM3** | HPM3 | **HAM3** | HPM3 | **HAM3** |
| MaxRM[a] (%) | 34.3248 | **13.9443** | 91.4172 | **18.4293** | 91.4095 | **18.4305** | 54.8021 | **13.9442** |
| MRE[a] (%) | 9.2866 | **7.0105** | 18.8254 | **7.6427** | 18.9200 | **7.6812** | 12.3098 | **8.1058** |
| RMSE[a] | 0.10545 | **0.04205** | 1676.594 | **668.548** | 0.02785 | **0.02092** | 0.1474 | **0.04429** |
| Sixth-order approximation | HPM6 | **HAM6** | HPM6 | **HAM6** | HPM6 | **HAM6** | HPM6 | **HAM6** |
| MaxRM (%) | 15.9332 | **2.4739** | 42.4349 | **4.4614** | 42.5597 | **4.4588** | 40.7392 | **7.5366** |
| MRE (%) | 3.1341 | **0.7891** | 7.0537 | **1.2610** | 7.0905 | **1.2674** | 5.2821 | **1.1598** |
| RMSE | 0.04259 | **0.00733** | 677.191 | **116.477** | 0.00895 | **0.00316** | 0.08464 | **0.01289** |

[a] MaxRE = $\max_{1 \leq i \leq N_P} \left|\frac{x_i^* - x_i}{x_i^*}\right|$; MRE = $\frac{1}{N_P}\sum_{i=1}^{N_P}\left|\frac{x_i^*-x_i}{x_i^*}\right|$; RMSE = $\sqrt{\frac{1}{N_P}\sum_{i=1}^{N_P}(x_i^*-x_i)^2}$, where $x_i^*$ is the exact solution (27) and $x_i$ is the analytic approximate solution (25). $N_P = 2000$.
[b] $x(\xi)$ is obtained from (29), assuming $\xi$ is a unknown variable when $t=1$.

*1) Verifying validity by fingerprints of memristor*

As seen in Fig. 2, the I-V and $R_M(x)$ curves, both based on the solved state variable $x_3(t)$ [from (3) and (6)], exhibit the characteristic of compressed hysteresis loops. The area of these loops continue to decrease monotonously with increasing frequency. These curves become straight lines as the frequency tends to infinity. This means the memristor converts to a linear resistor so its nonlinear characteristics disappear. The above results indicate that the HAM-based model conforms to the classical fingerprints of the memristor [40].

*2) Comparisons with HPM model and exact solution based model*

We can draw a conclusion from Figs. 3(a)–3(c) that the HAM-based model has faster convergence (which is identical to higher approximation accuracy under same $N$ [33]) and slower convergence errors than the HPM-based model by comparing with the exact solution (27) based model. To quantify the convergence error, defined by the difference between an analytic approximate solution and an exact or numerical solution, we performed the convergence error analyses by Maximum Relative Error (MaxRE), Mean Relative Error (MRE), and Root Mean Square Error (RMSE), respectively. Table IV shows that all the convergence errors of the HAM-based models are smaller than that of the HPM-based models, which is also an evidence of the conclusion in another aspect.

*3) Quantitative analysis of convergence*

To further investigate the difference of the convergence performance between the HAM-based model and the HPM-based model, we performed the quantitative analysis of the convergence, where the following intuitive and simple parameter, ratio $\beta$ [41], was used

$$\beta = \frac{\int_\Gamma x_{m+1}^2(t)dt}{\int_\Gamma x_m^2(t)dt} \quad (33)$$

where smaller $\beta$ corresponds to faster convergence.

Convergence analysis in Fig. 4 also demonstrates that the HAM-based model based on our methodology has faster convergence than the HPM-based model, because $\beta$ approaching to 0 monotonously as the approximation order $N$ increases. In contrast, the ratios of the HPM model at $N = 3$ and $N = 6$ are much larger than 1 (indicating slow convergence), and the ratios oscillate unexpectedly, which has an adverse impact on convergence [41].

*B. Highly nonlinearity*

Clearly, for the case of highly nonlinearity using J-window with $P = 2$ [35] and the associated governing equation

$$\frac{dx(t)}{dt} = \mu_V \frac{R_{ON}A}{D^2} I_M(t)(-16x^4 + 32x^3 - 24x^2 + 8x) \quad (34)$$

only a numerical solution of the state variable $x(t)$, but not an exact solution, can be reached. Note the highly nonlinearity here refers to the high-order nonlinear term in the governing equation of memristor, for example, the exponential term $x^4$ at the right side of (33).

To investigate the convergence of the different types of analytic models (based on HAM and HPM), we performed the following verifications:

*1) Comparisons with HPM model and numerical solution based model*

Fig. 5(a) shows the comparisons of the HAM-based and HPM-based models with the numerical solution based model, demonstrating that the HAM-based model offers faster conver-gence. Specifically, the HAM-based solutions of state variable $x(t)$ are in line with the numerical solutions (even the worst region in the inset). Nevertheless, it should be noted that all the HPM solutions, regardless of HPM3 or HPM6, exceed the physical boundaries ($x = 1$) in some regions. This violates the definition of memristor and may results in nonconvergence. In addition, increasing $N$ from 3 to 6 surprisingly increases convergence errors, which does not follow the general rule of the HPM-based model [30]. These observations are also confirmed by the convergence error analyses in Table V.

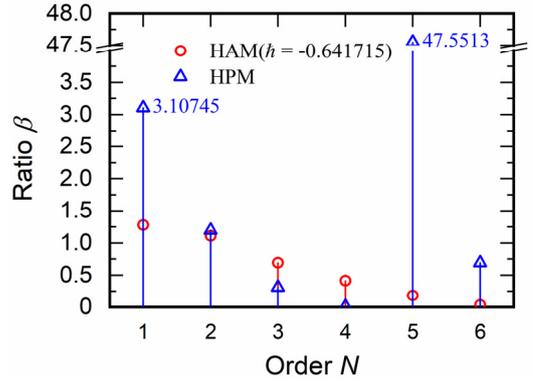

Fig. 4. Convergence analyses of the obtained HAM model and the HPM model [29]. $\beta$ is derived from (33). The J-window [35] with $P = 1$ is adopted. For simplicity, the interval of integration is limited to [0, 2000] that is identical to Table III. $\hbar$ of HAM3 is taken from Table III as an example, and the similar analyses can be done with different approximation orders. The simulation parameters of memristor and excitation are given in Table I. Note the ratios of HPM at $N = 1$ and $N = 6$ are 3.10745 and 47.5513, respectively.

*2) Effects of convergence errors on other parameters*

The state variable $x(t)$ is the core of the memristor that controls the other physical parameters, such as memristance (2) and voltage (3). Therefore, the convergence errors of the state variable $x(t)$ caused by the highly nonlinearity may transmit to these parameters. As shown in Figs. 5(b) and 5(c), both $R_M(x)$ and $V_M(x)$ calculated by the HPM-based model also exceed the practical physical boundaries.

In summary, the conclusions from the above observations can be reached:
1) The highly nonlinearity is a potential source of the deterioration of convergence.
2) The HAM-based methodology is more valid (showing faster convergence) for memristor modeling in the case of highly nonlinearity than the HPM-based ones.

The effect of the highly nonlinearity on convergence will be further explained in Section VI-C.

*C. Large $\xi$*

As can be seen from (28), $\xi$ represents the parameters of both the device and the input, so variations of $\xi$ directly affect the nonlinear term in the governing equation (34) and, thus, the convergence. Fig. 3(d) and Fig. 5(d) show $x-\xi$ analyses, demonstrating the HAM-based and the HPM-based models are both in well accordance with the exact and numerical solutions

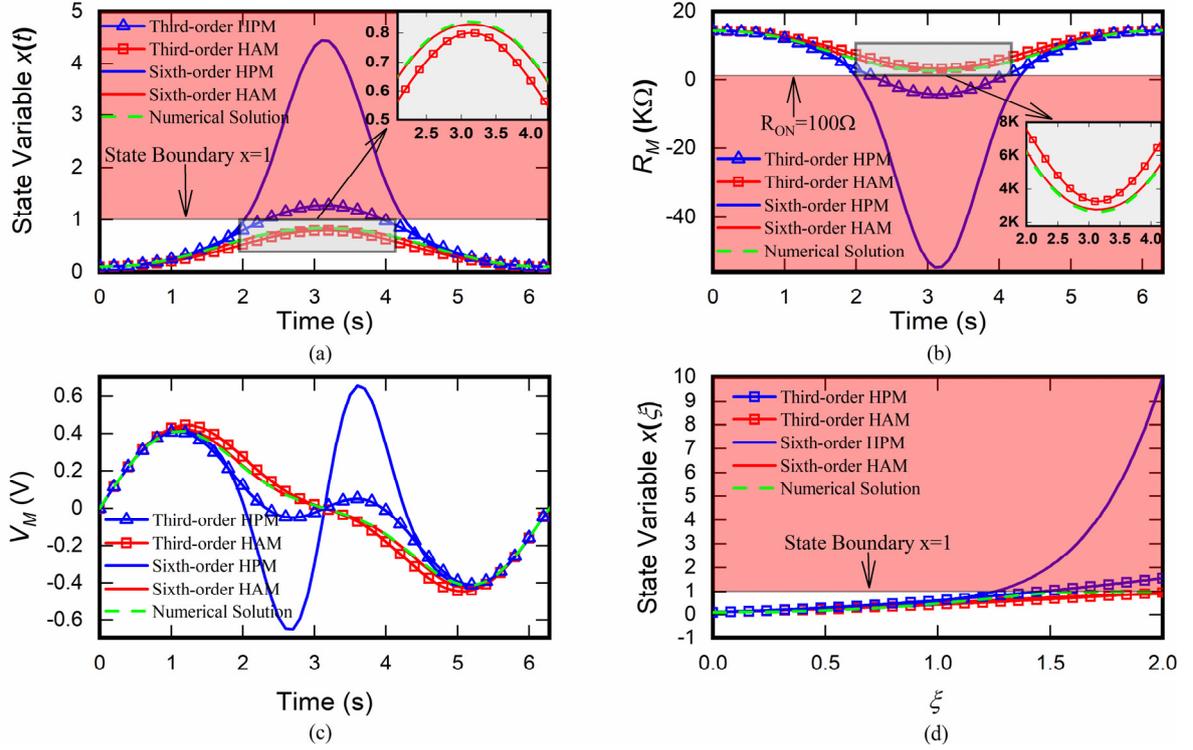

Fig. 5. Comparisons of the obtained HAM model and the HPM model [29] for the case of highly nonlinearity. (a)–(c) the evolution of $x(t)$, $R_M$, and $V_M$. (d) the $x-\xi$ curve when $t=1$. The J-window [35] with $P=2$ is adopted. $\hbar$ is given in Table III. All the other simulation configurations and parameters are the same as those in Fig. 3. Note that all the HPM solutions exceed the physical boundary (red rectangles indicate invalid area).

TABLE V
CONVERGENCE ERROR COMPARISONS OF THE HAM AND HPM MODELS WITH THE NUMERICAL SOLUTION BASED MODEL
WHEN USING J-WINDOW WITH $P=2$

| Key physical parameters | State variable $x(t)$ | | $R_M(t)$ | | $V_M(t)$ | | State variable $x(\xi)$ | |
|---|---|---|---|---|---|---|---|---|
| Third-order approximation | HPM3 | **HAM3** | HPM3 | **HAM3** | HPM3 | **HAM3** | HPM3 | **HAM3** |
| MaxRM (%) | 52.0878 | **18.7892** | 263.2321 | **29.3664** | 263.2322 | **29.3664** | 71.8159 | **43.0022** |
| MRE (%) | 18.6912 | **10.4382** | 58.5924 | **13.8757** | 58.8854 | **13.9451** | 36.6320 | **21.7442** |
| RMSE | 0.20733 | **0.05801** | 3296.612 | **922.273** | 0.06596 | **0.03120** | 0.19521 | **0.13519** |
| Sixth-order approximation | HPM6 | **HAM6** | HPM6 | **HAM6** | HPM6 | **HAM6** | HPM6 | **HAM6** |
| MaxRM (%) | 430.1029 | **6.9419** | 2173.5774 | **5.7644** | 2173.5781 | **5.7644** | 900.8238 | **57.2014** |
| MRE (%) | 90.7577 | **2.7459** | 360.4702 | **2.3311** | 362.2725 | **2.3427** | 139.1590 | **24.3999** |
| RMSE | 1.36709 | **0.01211** | 21736.769 | **192.561** | 0.30539 | **0.00689** | 2.38891 | **0.11002** |

* The calculations are the same as those in Table IV except $x_i^*$ is the numerical solution.

based models when $\xi < 1$. However, when $\xi > 1$, unlike HAM-based ones, the HPM-based models start deviating from the exact and numerical solutions based models. Moreover, the deviations become larger as $\xi$ increases, even exceeding the valid state boundary ($x=1$) in the case of highly nonlinearity as displayed in Fig. 5(d), which means nonconvergence. These limitations are resulted by the fact that the HPM-based model is more valid for the weakly nonlinear equations with small perturbations, but the large $\xi$ here is essentially a large perturbation [32], [33].

Following special emphases, based on the above analyses, can be reached:

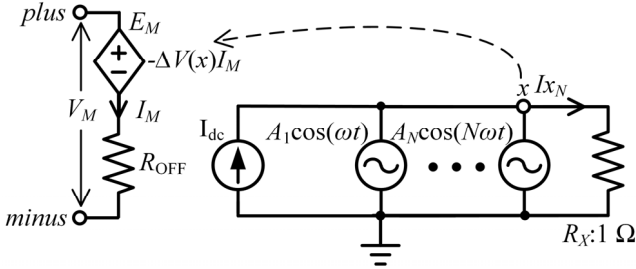

Fig. 6. Structure of the HAM-based Spice subcircuit.

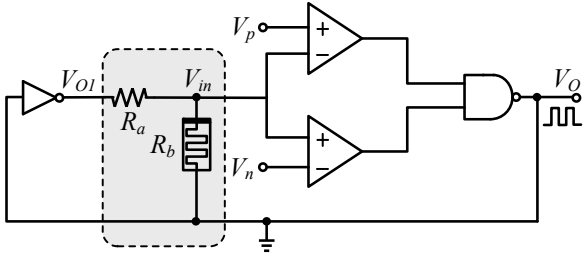

Fig. 7. Memristor-based Relaxation Oscillator.

1) Large $\xi$ has a negative effect on the convergence.
2) The HAM-based methodology is more valid (showing faster convergence) for memristor modeling in the case of large $\xi$ than the HPM-based ones.

The effect of the larger $\xi$ on the convergence will be further explained in Section VI-C.

## V. SPICE SUBCIRCUIT AND APPLICATION

A Spice subcircuit is built in this section for integrating the obtained HAM-based model into EDA tools. Then we applied this subcircuit to a Relaxation Oscillator, and compared Spectre simulation with theoretical calculation to verify the applicability of the established subcircuit. Finally, we verified its scalability on different simulation platforms.

### A. Spice subcircuit

The HAM-based Spice subcircuit is shown in Fig. 6, which is composed of a voltage controlled voltage source, a DC current source and $N$ AC current sources. $I_M$ is the input current and $V_M$ is the voltage across the memristor. $R_{\text{OFF}}$ is connected in series with the voltage-controlled current source $E_M$ [its port voltage is $x(R_{\text{OFF}} - R_{\text{ON}})$] to denote the memristance, which is in line with (3). The state variable $x$ is represented by node voltage $V(x)$ equal to $Ix_N R_x$. $R_x$ is an auxiliary resistor. The current $Ix_N$ consists of a DC current source $I_{\text{dc}}$ and $N$ AC current sources $A_N \cos(N\omega t)$, which correspond to the DC term (including the initial solution $x_0$) and the first-order fundamental to $N$th-order harmonic terms in (29), respectively. $A_N$ represents the magnitude of $N$th-order harmonic term.

Note that, our subcircuit, differs from traditional ones [24]–[26], did not adopt the method using the excitation current to charge and discharge an auxiliary capacitor for simulating the evolution of state variable. This method is a kind of numerical integration essentially, which may cause truncation and rounding errors [22].

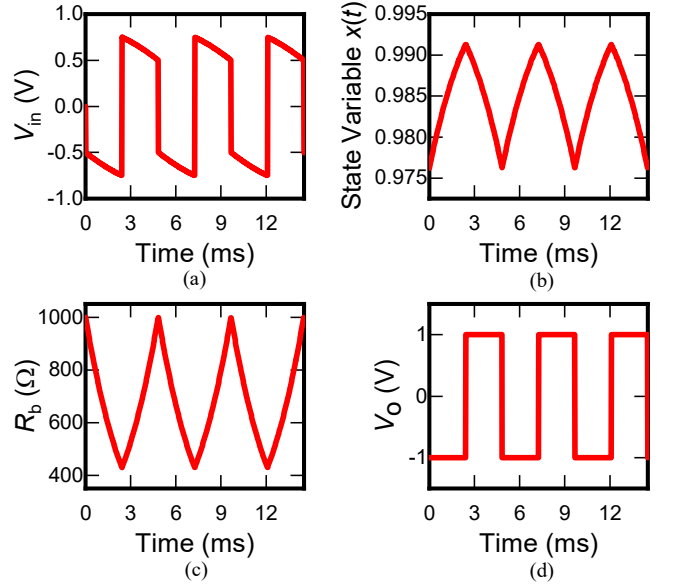

Fig. 8. Spectre simulations of the memristor-based Relaxation Oscillator. $R_a = 38$ K$\Omega$, $V_p = 0.5$ V, $V_n = -0.75$ V, $V_{ol} = -1$ V, $V_{oh} = 1$ V, and approximate order $N = 3$. The J-window [35] with $P = 2$ is adopted. The simulation parameters are given in Table I except: $R_{OFF} = 38$ K$\Omega$ and $x_0 = 0.97625$.

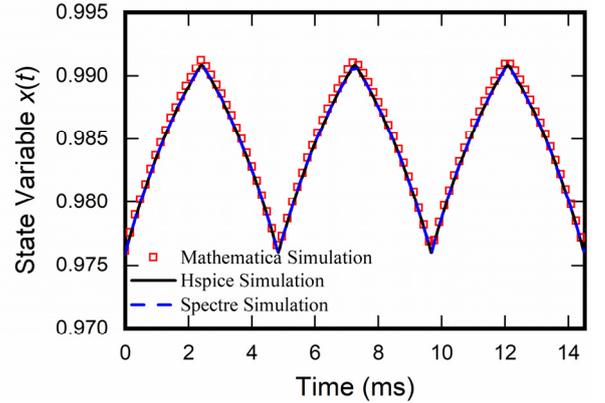

Fig. 9. Simulated state variables $x(t)$ of the memristor in the oscillator with various platforms. All the simulation configurations and parameters are the same as those in Fig. 8.

### B. Application

A test circuit—Relaxation Oscillator [42] was adopted for the verification of the applicability of the proposed subcircuit.

The memristor-based Relaxation Oscillator is shown in Fig. 7, where the RC circuit in traditional oscillator is replaced by the memristor $R_b$ which simulates the charge and discharge effects of the relaxation capacitor. $R_b$ decreases or increases when $V_{in}$ begins rising to $V_p$ or falling to $V_n$. Two comparators are used to guarantee the memristance varies between two val-ues restricted by $V_p$ and $V_n$.

TABLE VI
COMPARISONS OF THE PROPOSED HAM-BASED MEMRISTOR MODELING METHODOLOGY WITH OTHERS

| Modeling methodology | Microscopic physical methodology [19, 20] | Compact methodology [24-26] | Phenomenological methodology [27, 28] | HPM-based Methodology | **HAM-based Methodology** |
|---|---|---|---|---|---|
| Convergence problems | Numerous | Few | Nonexistent | Slow convergence | **Nonexistent** |
| Analyticity | No | Partially | Yes | Yes | **Yes** |
| Optimization of $\hbar$ | N/A | N/A | N/A | No | **Yes** |
| Accuracy of approximation for practical memristors | Highest | Low | Moderate | Limited | **Sufficient** |
| Suitable for physical mechanism with highly nonlinearity | Yes | No | Yes | No | **Yes** |
| Valid range of $\xi$ | N/A | N/A | N/A | Only mall | **Small/Large** |
| Scalability to other physical-based models | No | Partially | No | Partially | **Yes** |

This memristor-based oscillator takes the advantage of the *memristance-modulation* characteristic to adjust the period and the duty cycle of the oscillating output. Its oscillation frequency $f_0$ is given as

$$f_0 = \frac{(V_p - V_{ol})^2(V_n - V_{oh})^2}{R_a^2(V_{oh} - V_{ol})} \times \frac{k_b}{(V_{oh} + V_{ol} - V_p - V_n)(V_p - V_n + V_{oh} - V_{ol})} \quad (35)$$

where $k_b = \mu_V R_{ON}(R_{OFF} - R_{ON})f[x(t)]/D^2$, $V_{ol}$ and $V_{oh}$ represent the low voltage-level and the high voltage-level of the system, respectively. Fig. 8 displays the transient simulation results of the Relaxation Oscillator. It can be seen that even with no RC circuit, the basic function of the oscillator is well realized as expected. The state variable $x(t)$ oscillates between 0.97625 and 0.9913, resulting in $R_b$ changing between 1 KΩ and 428 Ω. The output $V_o$ is a square waveform with a frequency of 226 Hz which coincides well with the theoretical calculation of (35), indicating the well applicability and the accuracy of the HAM-based memristor subcircuit.

### C. Verification with different simulation platforms

In order to address the HAM-based model is suitable for various simulation platforms, we made comparisons among Mathematica (based on the analytic approximate solution), Hspice (based on the Spice subcircuit in Fig. 6), and Cadence Spectre simulations (based on a Verilog-A module). These simulation results displayed in Fig. 9 demonstrate the excellent consistency with each other.

## VI. DISCUSSION

To overcome the problems of nonconvergence in traditional nonanalytic models and slow convergence in current analytic models, a novel analytic memristor modeling methodology based on the HAM is proposed in this work. The summary of this methodology is shown in Table VI with detailed explanations as follows:

### A. Comparisons with nonanalytic modeling methodologies

*1) Convergence*

The evolving state variable $x(t)$ is described in the form of analytic approximate solution (19), avoiding the truncation and rounding errors caused by the numerical integration methods, commonly used in nonanalytic modeling methodologies [22]. That is the reason why the state variable $x(t)$ is convergent in our HAM-based methodology.

*2) Analyticity*

Owing to the complex physical properties of memristor, it is theoretically impossible to obtain a fully analytic solution. Different from traditional nonanalytic ones, the solution of the state variable $x(t)$ in HAM-based methodology is approximate analytic because it is represented by the homotopy-series (25) [31]–[33]. Due to this analyticity, the model produced by the methodology has the advantages of closed-form expression like (29) and symbolic computation. Therefore, the obtained state variable $x(t)$ could easily be employed to solve the analytic solutions of other parameters [see Figs. 3(b)–(c) and Figs. 5(b)–(c)].

### B. Comparisons with analytic modeling methodologies

*1) Convergence*

The convergence-control parameter $\hbar$ greatly affects the convergence of solution [39]. Thus, by solving the minimum value of $E_M(\hbar)$, the optimization of $\hbar$ (16) helps the analytic approximate solution quickly converge to the exact solution [see Fig. 3(a), Fig. 4, and Fig. 5(a)]. As a result, the approximation order $N$ needed to meet the specific convergence requirements is reduced.

*2) Approximation accuracy*

In general, fast convergence corresponds to high accuracy in device models [21]. Therefore, the model produced by the HAM-based methodology has sufficient approximation accuracy even in the case of low approximation order $N = 3$ (lower $N$ means lower accuracy), which is verified by the convergence error analyses (see Table IV and Table V) and the comparisons with the exact solution and the HPM based models [see Figs. 3(a)–(c) and Figs. 5(a)–(c)].

TABLE VII
CLASSIFICATIONS, MECHANISMS, AND GOVERNING EQUATIONS OF SOME CURRENT MEMORY NANODEVICES

| Classifications | Conductive bridging RAM (CBRAM) [43] | Valence change mechanism RRAM (VCM-RRAM) [44] | Two-dimension nanomaterials based resistive device [45] | Simmons tunneling based memristive device [46] |
|---|---|---|---|---|
| Physical mechanisms | Active metal based filaments undergo Redox reactions in electrolytes | Generation, recombination, and hopping of oxygen-vacancies change valence of local metal ions | Vacancies migration and mobile ions penetration in a stack | Modulation of a tunneling gap |
| Simplifications of right side of the governing equations | $\sinh(x)$ | $\sinh(x^3)$ | $x + e^x$ | $e^{(x+e^x)}$ |
| Physical significances of the state variable $x$ | Length of conductive filament | Gap between tip of filament and top electrode | Total number of conductive filaments in a constriction region | Simmons tunnel barrier width |

*C. Highly nonlinearity and large ξ*

The highly nonlinearity in governing equation and large ξ in solution expression may be due to the complex physical mechanisms of memristors. For example, the motions of cations and anions are simultaneously affected by different factors, such as internal electric field, Joule heat, chemical energy, and activation energy. In these cases, the HPM-based analytic model has slower convergence and larger approximation errors than the case of weakly nonlinearity (Section IV-A), even nonconvergence happens (see Fig. 5(d) and Table V). Although we only verified the HPM-based analytic models, the similar analysis can be also applied to other analytic models because these cases are common phenomena in memristors. Therefore, the highly nonlinearity and large ξ can be revealed as the potential sources of the slow convergence in analytic memristor models.

In contrast, by employing the convergence-control parameter $\hbar$, which guarantees the convergence of the approximation, the model produced by the HAM-based methodology has faster convergence and smaller approximation errors than the HPM-based model in these cases (see Fig. 5(d) and Table V). These observations further indicate that the HAM-based methodology is more suitable for memristor modeling in the case of highly nonlinearity and large ξ than the HPM-based methodology.

*D. Scalability*

Although the physical mechanisms differ significantly in various memory devices, most of them can be represented by the governing equations (for example, (1) of the HP memristor) that are essentially nonlinear differential equations [21]. The differences of these equations are the definitions of state variables, the range of physical parameters, and the nonlinear terms, as shown in Table VII. In theory, the HAM-based modeling methodology could be applied to various memory devices, because HAM is a general mathematical method for solving nonlinear differential equations [31]–[33]. In addition, the model obtained from this methodology is scalable for various simulation platforms, as can be seen in Fig. 9.

VII. CONCLUSION

In summary, we have presented a novel analytic modeling methodology for memristor. To the best of the authors' knowledge, this is the first report that HAM is applied to memristor modeling. The HAM-based methodology is characterized by the fast convergence under the premise of sufficient accuracy and has the advantages of high scalability. Based on this methodology, we have revealed that highly nonlinearity and large ξ are the potential sources of the slow convergence in analytic memristor models.

REFERENCES

[1] Q. Luo et al., "8-Layers 3D vertical RRAM with excellent scalability towards storage class memory applications," in *IEDM Tech. Dig.*, San Francisco, CA, USA, Dec. 2017, pp. 2.7.1–2.7.4.
[2] Q. Luo et al., "Self-Rectifying and Forming-Free Resistive-Switching Device for Embedded Memory Application," *IEEE Electron Device Lett.*, vol. 39, no. 5, pp. 664–667, May. 2018.
[3] J. M. Portal et al., "Design and Simulation of a 128 kb Embedded Non-volatile Memory Based on a Hybrid RRAM (HfO$_2$)/28 nm FDSOI CMOS Technology," *IEEE Trans. Nanotechnol.*, vol. 16, no. 4, pp. 677–686, Jul. 2017.
[4] H. Abbas et al., "A memristor crossbar array of titanium oxide for non-volatile memory and neuromorphic applications," *Semicond. Sci. Technol.*, vol. 32, no. 6, p. 7, Jun. 2017.
[5] Y. van de Burgt et al., "A non-volatile organic electrochemical device as a low-voltage artificial synapse for neuromorphic computing," *Nat. Mater.*, vol. 16, no. 4, pp. 414–418, Apr. 2017.
[6] P. M. Sheridan, F. X. Cai, C. Du, W. Ma, Z. Y. Zhang, and W. D. Lu, "Sparse coding with memristor networks," *Nat. Nanotechnol.*, vol. 12, no. 8, pp. 784–789, Aug. 2017.
[7] Z. R. Wang et al., "Memristors with diffusive dynamics as synaptic emulators for neuromorphic computing," *Nat. Mater.*, vol. 16, no. 1, pp. 101–108, Jan. 2017.
[8] C. Li et al., "Efficient and self-adaptive in-situ learning in multilayer memristor neural networks," *Nat. Commun.*, vol. 9, p. 8, Jun. 2018.
[9] V. Milo, D. Ielmini, and E. Chicca, "Attractor networks and associative memories with STDP learning in RRAM synapses," in *IEDM Tech. Dig.*, San Francisco, CA, USA, Dec. 2017, pp. 11.2.1–11.2.4.

[10] W. H. Chen et al., "A 65nm 1Mb nonvolatile computing-in-memory ReRAM macro with sub-16ns multiply-and-accumulate for binary DNN AI edge processors," in *IEEE ISSCC Dig. Tech. Paper*, San Francisco, CA, USA, Feb. 2018, pp. 494–496.

[11] S. Ding, Z. Wang, Z. Huang, and H. Zhang, "Novel Switching Jumps Dependent Exponential Synchronization Criteria for Memristor-Based Neural Networks," *Neural Process. Lett.*, vol. 45, no. 1, pp. 15–28, Feb. 2016.

[12] J. D. Cao and R. X. Li, "Fixed-time synchronization of delayed memristor-based recurrent neural networks," *Sci. China-Inf. Sci.*, vol. 60, no. 3, p. 15, Mar. 2017.

[13] B. J. Choi et al., "High-Speed and Low-Energy Nitride Memristors," *Adv. Funct. Mater.*, vol. 26, no. 29, pp. 5290–5296, Aug. 2016.

[14] X. Xiaoxin et al., "Fully CMOS compatible 3D vertical RRAM with self-aligned self-selective cell enabling sub-5nm scaling," in *Proc. Symp. VLSI Technol.*, Honolulu, HI, USA, Jun. 2016, pp. 1–2.

[15] P. Bousoulas, I. Michelakaki, E. Skotadis, M. Tsigkourakos, and D. Tsoukalas, "Low-Power Forming Free $TiO_{2-x}$/$HfO_{2-y}$/$TiO_{2-x}$-Trilayer RRAM Devices Exhibiting Synaptic Property Characteristics," *IEEE Trans. Electron Devices*, vol. 64, no. 8, pp. 3151–3158, Aug. 2017.

[16] J. T. Zhou, F. X. Cai, Q. W. Wang, B. Chen, S. Gaba, and W. D. Lu, "Very Low-Programming-Current RRAM With Self-Rectifying Characteristics," *IEEE Electron Device Lett.*, vol. 37, no. 4, pp. 404–407, Apr. 2016.

[17] D. B. Strukov, G. S. Snider, D. R. Stewart, and R. S. Williams, "The missing memristor found," *Nature*, vol. 453, pp. 80–83, 2008.

[18] International technology roadmap for semiconductors (ITRS). [Online]. Available: http://www.itrs2.net/itrs-reports.html

[19] F. Pan, S. Yin, and V. Subramanian, "A Detailed Study of the Forming Stage of an Electrochemical Resistive Switching Memory by KMC Simulation," *IEEE Electron Device Lett.*, vol. 32, no. 7, pp. 949–951, Jul. 2011.

[20] S. Yu, G. Ximeng, and H. S. P. Wong, "On the stochastic nature of resistive switching in metal oxide RRAM: Physical modeling, monte carlo simulation, and experimental characterization," in *2011 International Electron Devices Meeting IEDM Tech. Dig.*, Washington, DC, USA, Dec. 2011, pp. 17.3.1–17.3.4.

[21] A. Ascoli, R. Tetzlaff, Z. Biolek, Z. Kolka, V. Biolkova, and D. Biolek, "The Art of Finding Accurate Memristor Model Solutions," *IEEE J EM SEL TOP C*, vol. 5, no. 2, pp. 133–142, Jun. 2015.

[22] I. Messaris, A. Serb, S. Stathopoulos, A. Khiat, S. Nikolaidis, and T. Prodromakis, "A Data-Driven Verilog-A ReRAM Model," *IEEE Trans. Comput.-Aided Des.Integr. Circuits Syst.*, p. 1, Jan. 2018.

[23] D. Biolek, Z. Kolka, V. Biolková, Z. Biolek, M. Potrebić, and D. Tošić, "Modeling and simulation of large memristive networks," *Int. J. Circuit Theory Appl.*, vol. 46, no. 1, pp. 50–65, Jan. 2018.

[24] S. Sangho, Z. Le, G. Weickhardt, C. Seongik, and K. Sung-Mo, "Compact Circuit Model and Hardware Emulation for Floating Memristor Devices," *IEEE Circuits Syst. Mag.*, vol. 13, no. 2, pp. 42–55, May. 2013.

[25] X. Wang, S. Li, H. Liu, and Z. Zeng, "A Compact Scheme of Reading and Writing for Memristor-Based Multi-Valued Memory," *IEEE Trans. Comput.-Aided Des.Integr. Circuits Syst.*, vol. 37, no. 7, pp. 1505–1509, Jul. 2017.

[26] R. Zhu, S. Chang, H. Wang, Q. Huang, J. He, and F. Yi, "A Versatile and Accurate Compact Model of Memristor With Equivalent Resistor Topology," *IEEE Electron Device Lett.*, vol. 38, no. 10, pp. 1367–1370, Oct. 2017.

[27] S. Yu and H. S. P. Wong, "A Phenomenological Model for the Reset Mechanism of Metal Oxide RRAM," *IEEE Electron Device Lett.*, vol. 31, no. 12, pp. 1455–1457, Dec. 2010.

[28] F. Merrikh Bayat, B. Hoskins, and D. B. Strukov, "Phenomenological modeling of memristive devices," *Appl. Phys. A-Mater. Sci. Process.*, vol. 118, no. 3, pp. 779–786, Mar. 2015.

[29] C. Hernández-Mejía, A. Sarmiento-Reyes, and H. Vázquez-Leal, "A Novel Modeling Methodology for Memristive Systems Using Homotopy Perturbation Methods," *Circuits Syst. Signal Process.*, vol. 36, no. 3, pp. 947–968, Mar. 2017.

[30] J. H. He, "A coupling method of a homotopy technique and a perturbation technique for non-linear problems," *Int. J. Non-Linear Mech.*, vol. 35, no. 1, pp. 37–43, Jan. 2000.

[31] S. J. Liao, "The proposed Homotopy Analysis Technique for the Solution of Nonlinear Problems," Ph.D. dissertation, Dept. School of Naval Architecture & Ocean Engineering, Shanghai Jiaotong Univ., Shanghai, China, 1992.

[32] S. J. Liao, "Systematic descriptions and related theorems," in *Homotopy Analysis Method in Nonlinear Differential Equations*, Heidelberg, Germany: Springer & Higher Education Press, 2012, ch. 4, sec. 1–6, pp. 135–187.

[33] S. J. Liao, "Chance and Challenge: A Brief Review of Homotopy Analysis Method," in *Advances in the Homotopy Analysis Method*, Singapore: World Scientific, 2013, ch. 1, sec. 2–3, pp. 103–160.

[34] Z. Biolek, D. Biolek, and V. Biolkova, "SPICE Model of Memristor with Nonlinear Dopant Drift," *Radioengineering*, vol. 18, no. 2, pp. 210–214, June. 2009.

[35] Y. N. Joglekar and S. J. Wolf, "The elusive memristor: properties of basic electrical circuits," *Eur. J. Phys.*, vol. 30, no. 4, pp. 661–675, May. 2009.

[36] T. Prodromakis, B. P. Peh, C. Papavassiliou, and C. Toumazou, "A Versatile Memristor Model With Nonlinear Dopant Kinetics," *IEEE Trans. Electron Devices*, vol. 58, no. 9, pp. 3099–3105, Sep. 2011.

[37] S. Kvatinsky, M. Ramadan, E. G. Friedman, and A. Kolodny, "VTEAM: A General Model for Voltage-Controlled Memristors," *EE Trans. Circuits Syst. II, Express Briefs*, vol. 62, no. 8, pp. 786–790, 2015.

[38] S. J. Liao, "Illustrative description," in *Beyond Perturbation: Introduction to the Homotopy Analysis Method*, Boca Raton, FL, USA: Chapman & Hall/CRC, 2003, ch. 2, sec. 3, pp. 11–13.

[39] S. J. Liao, "An optimal homotopy-analysis approach for strongly nonlinear differential equations," *Commun. Nonlinear Sci. Numer.*, vol. 15, no. 8, pp. 2003–2016, Aug. 2010.

[40] S. P. Adhikari, M. P. Sah, H. Kim, and L. O. Chua, "Three Fingerprints of Memristor," *IEEE Trans. Circuits Syst. I, Reg. Papers*, vol. 60, no. 11, pp. 3008–3021, Nov. 2013.

[41] M. Turkyilmazoglu, "An effective approach for evaluation of the optimal convergence control parameter in the homotopy analysis method," *Filomat*, vol. 30, no. 6, pp. 1633–1650, 2016.

[42] M. E. Fouda, M. A. Khatib, A. G. Mosad, and A. G. Radwan, "Generalized Analysis of Symmetric and Asymmetric Memristive Two-Gate Relaxation Oscillators," *IEEE Trans. Circuits Syst. I, Reg. Papers*, vol. 60, no. 10, pp. 2701–2708, Oct. 2013.

[43] Y. D. Zhao et al., "A physics-based compact model for material- and operation-oriented switching behaviors of CBRAM," in *2016 IEEE International Electron Devices Meeting (IEDM) IEDM Tech. Dig.*, San Francisco, CA, USA, Dec. 2016, pp. 7.6.1–7.6.4.

[44] H. Li, T. F. Wu, S. Mitra, and H. S. P. Wong, "Resistive RAM-Centric Computing: Design and Modeling Methodology," *IEEE Trans. Circuits Syst. I, Reg. Papers*, vol. 64, no. 9, pp. 2263–2273, Sep. 2017.

[45] C. Pan et al., "Model for multi-filamentary conduction in graphene/hexagonal-boron-nitride/graphene based resistive switching devices," *2D Mater.*, vol. 4, no. 2, p. 025099, May. 2017.

[46] M. D. Pickett et al., "Switching dynamics in titanium dioxide memristive devices," *J. Appl. Phys.*, vol. 106, no. 7, p. 074508, Oct. 2009.